\documentstyle[amsfonts,prl,aps,preprint,tighten]{revtex}

\newbox\rotbox

\begin{document}

\preprint{\vbox{\noindent{\it Submitted to Phys.\ Lett.\ B}
                          \hfill UW-DOE/ER/40427-2-N96\\ 
                    \null \hfill TRI-PP-96-11 \\
}}

\draft

\title{\huge Background-Field Formalism\\ in\\ Nonperturbative QCD}
\author{\sc Matthias Burkardt,$^{\rm
1}$\footnote{
E-mail:~burkardt@nmsu.edu ~$\bullet$~
Telephone: (505) 646--1928 ~$\bullet$~ Fax: (505) 646--1934}
Derek B. Leinweber$^{\rm 2,3}$\footnote{
E-mail:~derek@phys.washington.edu ~$\bullet$~ Telephone:
(206)~616--1447 ~$\bullet$~ Fax: (206)~685--0635 \hfill\break
\null\quad\quad WWW:~http://www.phys.washington.edu/$\sim$derek/}
and Xuemin Jin$^{\rm 3}$\footnote{
Address after September 1, 1996: Center for Theoretical Physics,
Laboratory for Nuclear Science and Department of Physics,
Massachusetts Institute of Technology, Cambridge, 
MA 02139} \\
\null} 

\address{$^{\rm 1}$Department of Physics, New Mexico State University,
                   Las Cruces, NM 88003--0001, USA\\
               and Institute for Nuclear Theory, Box 351550, 
                   University of Washington, Seattle WA 98195, USA} 

\address{$^{\rm 2}$Department of Physics, Box 351560, 
                   University of Washington,
                   Seattle, WA 98195, USA}

\address{$^{\rm 3}$TRIUMF, 4004 Wesbrook Mall, 
                   Vancouver, B.C,
                   V6T 2A3, Canada} 
\maketitle

\begin{abstract}
The background-field formalism is used extensively in fundamental
approaches to QCD to explore hadronic matrix elements of various
currents.  While the lattice QCD approach is formulated in the
fully-interacting Hilbert space, which includes both QCD and
background field interactions, the QCD sum-rule formalism is
traditionally developed in the pure-QCD Hilbert space.  The latter
approach encounters difficulties with excited state contaminations
which are not exponentially suppressed and requires extrapolations to
isolate the desired physics.  Proponents of the pure-QCD Hilbert space
formalism used in the QCD sum-rule approach have criticized the
lattice QCD approach as neglecting important physics.  In this letter,
the equivalence of the two approaches is established and the flaws in
the former criticisms are resolved.  Finally, the application of the
fully-interacting Hilbert space formalism to the study of
electromagnetic polarizabilities is outlined.

\vspace{24pt}

\end{abstract}
\newpage
\narrowtext

   The background-field formalism of field theoretic approaches to
hadron phenomenology has been used extensively to resolve the static
properties of hadrons.  For example, early lattice QCD calculations of
baryon magnetic moments \cite{bernard82,martinelli82} introduced a
background magnetic field into the gauge field configurations.  The
magnetic moments were extracted by calculating the energy shift of the
ground state in the presence of the background field which is
proportional to $\bbox{\mu} \cdot \bbox{B}$. The QCD sum-rule approach
has also exploited the background-field formalism to study hadron
properties such as magnetic moments \cite{ioffe84,chiu86}.  However,
the description for propagating hadrons in the presence of the
background field is handled differently in these two approaches.

   In the lattice QCD approach, the effects of the background field
are treated in a nonperturbative sense.  The propagating quarks
experience the background field at each link throughout the lattice
volume such that complex motion such as the zero-point orbital motion
can evolve.  When inserting a complete set of hadron states to
describe the propagator, the Hilbert space of the full Hamiltonian,
including both QCD and background-field effects, arises naturally.

   On the other hand, only the linear response to the background field
is considered in the traditional QCD sum-rule approach. In this case
the effects of the background field are treated perturbatively at the
phenomenological level.  It is worth emphasizing that the
background-field QCD sum-rule approach, usually adopted in the
literature, is identical to the QCD sum-rule approach based on direct
three-point function considerations, and the introduction of the
background field is only for bookkeeping.  In the QCD sum-rule
background-field formalism, one can select the Hilbert space of either
the full Hamiltonian or the normal QCD Hamiltonian.  However, as we
will show, one ultimately needs to resort to the Hilbert space of the
normal QCD Hamiltonian in order to extract the hadron properties of
interest.

   A well-known problem that arises in the QCD sum-rule
background-field formalism is the unwanted physics associated with
transitions from the ground state hadron to excited states. The
contributions of these transitions are not exponentially suppressed
(after Borel transformation) relative to the ground state contribution
which contains the ground state property of interest.  This
drawback is general to any QCD sum-rule calculation based on
three-point functions.

  In principle, there are infinitely many transition terms as there
are infinitely many excited states, and they should be included in the
QCD sum-rule calculations. Fortunately, the polynomial (in Borel mass)
behavior of the ground state signal is different from those of the
transition terms. The usual approximation is to introduce a new
unknown phenomenological parameter (independent of Borel mass)
accounting for the sum over the contributions from all the transitions
between the ground state and the excited states.  This new
parameter is extracted from the sum rules, along with the ground state
property.  The hope is that a sufficiently large valid Borel regime
exists such that the interplay of the two parameters may be resolved.
While this approximation has been used in earlier QCD sum-rule
studies, it has been noticed recently that the parameter representing
the transitions is in general dependent on the Borel mass, which may
have a sizable impact on the extracted ground state properties
\cite{ioffe95,jin95}.

   In Ref. \cite{ioffe84} Ioffe and Smilga argued that the transition
terms are also problematic in lattice QCD calculations employing the
background-field approach.  They claim that the transition
contributions were overlooked in Ref. \cite{bernard82,martinelli82}
and may be as large as 1/3 of the signal associated with the ground
state magnetic moment.

   Once the sequential source technique \cite{bernard86} was
established for calculating the propagators encountered in three-point
functions, the consideration of hadronic matrix elements became the
method of choice for determining hadron properties such as
electromagnetic form factors
\cite{draper90,leinweber91,leinweber92b,leinweber93e}.  However, given
the difficulties associated with calculating four-point functions in
lattice field theory\cite{negele}, the background field formalism may
be the best way to make contact with nucleon electromagnetic
polarizabilities and other aspects of the Compton scattering program
of TJNAF.

   In this letter we will examine the relationship between the two
apparently different approaches of lattice QCD and QCD sum rules.  We
will illustrate how the problematic contaminations of the transition
terms encountered in the QCD sum-rule formalism are specific to the
QCD sum-rule approach. There, only the linear response term in the
background field is considered and hence the Hilbert space of the
normal QCD Hamiltonian is employed.  Moreover, we will illustrate that
to first order in the background field strength the two approaches are
indeed equivalent, and that the problematic excited state
contaminations simply reflect the unfortunate but necessary resort to
a Hilbert space that is not the fully-interacting Hilbert space.


   In anticipation of applying the background-field formalism to the
study of polarizabilities in lattice field theory, we will formulate
our discussion in Euclidean space-time.  The results are easily
transformed to the QCD sum-rule formalism.

   Consider the change in the two-point function due to the presence
of a background field.  If $\delta H$ is the change in the Hamiltonian
reflecting the background field effects, then to leading order in
$\delta H$ \cite{leinweber91} we have
\begin{equation}
\delta G(t_2;\bbox{p,q}) = \int_0^{t_2} dt_1 \, 
   \int d^3x_2 \, d^3x_1 \,
   e^{-i \bbox{p} \cdot \bbox{x_2} } 
   e^{+ i \bbox{q} \cdot \bbox{x_1} } 
   \langle \Omega | {\rm T} \left [ \chi(x_2) \, \delta H(x_1) \,
   \overline\chi(0) \right ] 
                  | \Omega \rangle \, .
\end{equation}
Here $\bbox{q}$ denotes the three-momentum of the interaction and
$\bbox{p}$ denotes the momentum of the final state.  $\chi$ is a
hadron interpolating field constructed from local quark and gluon
operators identifying the spin and isospin of the hadron under
investigation.  Since intermediate states propagate on shell, a
variety of momentum transfers may be studied.  However, our focus here
is on zero momentum transfer which is easily accessed via
$\bbox{p}=\bbox{q}=0$.  Hence we focus on 
\begin{equation}
\delta G(T) = \int_0^{T} dt \, 
      \langle \Omega | {\rm T} \left [ \chi(T) \, \delta H(t) \, 
      \overline\chi(0) \right ] 
                  | \Omega \rangle \, ,
\end{equation}
where $T=t_2$, $t=t_1$ and the integrals over spatial coordinates are
implicit.

To make contact with the QCD sum-rule formalism, we proceed by
inserting a complete set of states in the normal QCD Hilbert space
denoted by $|N\rangle$ and $|N^*\rangle$ indicating ground and
excited states, respectively. 
\begin{eqnarray}
\delta G(T) &=& \int_0^T dt
\left[ \langle \Omega|\chi |N \rangle 
e^{-M_N(T-t)} 
\langle N|\delta H|N\rangle 
e^{-M_Nt}
\langle N|\overline\chi |\Omega \rangle \right. 
\nonumber\\
& &
+ \sum_{N^*\neq N} \langle \Omega|\chi |N^* \rangle 
e^{-M_{N^*}(T-t)} 
\langle N^*|\delta H|N\rangle 
e^{-M_{N}t}
\langle N|\overline\chi |\Omega \rangle 
\nonumber\\
& &
+ \sum_{N^*\neq N} \left. \langle \Omega|\chi |N \rangle 
e^{-M_{N}(T-t)} 
\langle N|\delta H|N^*\rangle 
e^{-M_{N^*} t}
\langle N^*|\overline\chi |\Omega \rangle \right ]
\nonumber\\
& &
+ \;\text{exponentially suppressed terms.}
\label{eq:g1}
\end{eqnarray}
Since $t$ is integrated from $0$ to $T$ the interaction Hamiltonian
$\delta H(t)$ can get arbitrarily close to the source
$(\overline\chi)$ or sink $(\chi)$.  Therefore, even as $T \rightarrow
\infty$, terms that include contributions from hadron resonances are
{\it not} exponentially suppressed compared to the ground state
property one is interested in.  In fact, explicit integration
yields
\begin{eqnarray}
\delta G(T) &=& 
\langle \Omega|\chi |N \rangle \,
\langle N|\delta H|N\rangle \, 
\langle N |\overline\chi |\Omega \rangle \,
T \, e^{-M_N T} \nonumber\\
& &+\sum_{N^*\neq N}
\Biggl \{ \Bigl [ \langle \Omega|\chi |N^* \rangle \, 
\langle N^*|\delta H|N\rangle \, \langle N|\overline\chi |\Omega
\rangle 
\nonumber\\
& &\qquad\qquad+  \langle \Omega|\chi |N \rangle \,
\langle N|\delta H|N^*\rangle \, \langle N^*|\overline\chi |\Omega
\rangle \Bigr ] \,
\frac{e^{-(M_{N^*}-M_N)T} - 1 }{M_N-M_{N^*}} \Biggr \} \, e^{-M_N T}
\nonumber\\
& &+\;\text{exponentially suppressed terms.}
\label{eq:g2}
\end{eqnarray}
The latter term containing the transition matrix elements essentially
falls off as $\exp(-M_N T)$, demonstrating that the transition
contribution is only power law suppressed in T.  This excited-state
contamination was first encountered in the QCD sum-rule approach.  The
usual approximation is to introduce a new fit parameter accounting for
the quantity in braces and use the different $T$ (or squared Borel
mass, $M^2$, in QCD sum rules) dependence to isolate the ground state
signal.  It is not clear how this treatment affects the predictive
ability of the QCD sum-rule approach, as a realistic uncertainty
analysis \cite{leinweber95d} has not been done.  Nevertheless, error
estimates of the standard 10\% are claimed.

   Eq.\ (\ref{eq:g2}) also shows how a naive evaluation of
background-field two-point functions in Euclidean lattice QCD may
contain large contaminations from transitions to the $N^*$ as well.
While lattices are large enough that exponentially suppressed terms
are negligible, this cannot be said about power law suppressed pieces.
We note however for clarity, that these problems are not encountered
in the standard treatment of three-point functions on the lattice.
There, the intermediate point $t$ is fixed at large values to ensure
exponential suppression of excited states prior to interaction with
the probing current.  The final time $T$ is taken to be $\gg t$ to
once again ensure isolation of the ground state signal following
interaction with the current.

Now consider the two-point function in the presence of the background
field to all orders in $\delta H$
\begin{equation}
G(T) = \langle \Omega | {\rm T} \left [ \chi(T) \overline\chi(0)
                                  \right ]  
       | \Omega \rangle \, .
\label{AllOrders}
\end{equation}
This is the approach adopted in background-field lattice QCD
calculations. If one denotes by $|N^\prime\rangle $, $|{N^*}^\prime
\rangle$ the exact eigenstates in the fully-interacting Hilbert space,
then
\begin{equation}
G(T) = 
\langle \Omega|\chi |N^\prime \rangle \, 
\langle N^\prime |\overline\chi | \Omega \rangle \,
e^{-M_{N^\prime}T}
\quad +\quad \text{exponentially suppressed terms}.
\label{eq:g3}
\end{equation}
In Eq.(\ref{eq:g3}) the transitions to excited states are hidden in
the implicit dependence of the basis states on the background field,
as reflected in the first order perturbation theory formula
\begin{equation}
|N^\prime \rangle = | N\rangle + \sum_{N^*\neq N} \frac{|N^*\rangle
\langle N^*|\delta H |N\rangle}{M_N-M_{N^*}}+ {\cal O}(\delta H^2) 
\, ,
\label{WaveExp}
\end{equation}
yielding the dependence of the ``coupling constants'' on the
background field
\begin{equation}
\langle \Omega|\chi|N^\prime \rangle = \langle \Omega|\chi| N\rangle +
\sum_{N^*\neq N} \frac{\langle \Omega|\chi|N^*\rangle
\langle N^*|\delta H |N\rangle}{M_N-M_{N^*}}+ {\cal O}(\delta H^2) 
\, .
\end{equation}
The equivalence of Eq.(\ref{eq:g2}) and Eq.(\ref{eq:g3}) can
then be verified by utilizing the perturbative expansion of the mass
\begin{equation}
M_{N'} = M_N + \langle N | \delta H | N \rangle + {\cal O}(\delta H^2)
\, ,
\label{MassExp}
\end{equation}
and expanding the exponential to leading order
\begin{equation}
e^{- \langle N | \delta H | N \rangle \, T} = 
1 - \langle N | \delta H | N \rangle \, T + {\cal O}(\delta H^2) \, .
\label{ExpExp}
\end{equation}

  We observe that the transition contaminations originate from
the response of the hadron wave function to the background field 
as shown in Eq.~(\ref{WaveExp}). This elementary observation clearly 
indicates how one may avoid the problem of contamination from 
excited states.  
%
%
First one calculates the two point function of Eq.\ (\ref{AllOrders})
in the presence of the background field.  Rather than introducing
pure-QCD parameters such as the nucleon mass or nucleon coupling
constants of the interpolators, one simply extracts the mass of the
ground state in the fully-interacting Hilbert space by fitting the
slope of the logarithm of the lattice two-point correlation function.
By determining the dependence of the ground state mass as a function
of background-field strength, the magnetic moment may be extracted
from the slope as $|\bbox{B}| \to 0$.  This has always been the
procedure used by lattice QCD practitioners in the background-field
formalism.

On the other hand, if one evaluates the two-point function only to
leading order in $\delta H$, as in the usual QCD sum-rule formalism,
the transition contamination will be unavoidable.  This can be seen by
expanding the exponential of (\ref{eq:g3}) to leading order
\begin{equation}
\delta G(T) = \langle \Omega|\chi |N^\prime \rangle \,
\langle N^\prime |\overline\chi | \Omega \rangle \,
\langle N | \delta H | N \rangle \, T \, e^{-M_N \, T} \, .
\end{equation}
It is impossible to resolve both the unknown couplings, $\langle
\Omega|\chi |N^\prime \rangle \, \langle N^\prime |\overline\chi |
\Omega \rangle$, and the matrix element of interest, $\langle N |
\delta H | N \rangle$.  Hence, one is forced to expand the fully-interacting
Hilbert space couplings in order to make contact with the normal QCD
Hilbert space couplings determined from other correlation functions.
In doing so, one must confront the contaminations of excited state
contributions which are not exponentially suppressed.

   Finally, it is tempting to adopt Eq.\ (\ref{eq:g3}) in the QCD
sum-rule formalism.  This, however, requires a new method to evaluate
the two-point function to all orders in $\delta H$.  Such a method is
not available to date. Nevertheless, it will be interesting to
determine the uncertainty associated with $\langle N | \delta H | N
\rangle$ in the QCD sum-rule approach.  It is well known that the
interpolating field couplings show much larger relative errors than
the exponential decay parameter such as the mass \cite{leinweber95d}.
The matrix element $\langle N | \delta H | N \rangle$ and the
couplings play very similar roles as illustrated in (\ref{eq:g2}).
Hence it is important to establish the true predictive ability of QCD
sum rules for hadron matrix elements.  Research in this direction is
in progress \cite{lee96b,hencken96}.

   Having established the viability of the background field formalism,
it is interesting to consider the possibility of determining second
order effects of the background field.  Electromagnetic
polarizabilities may be extracted from (\ref{eq:g3}) by simply
expanding the extracted mass to second order in $\delta H$.  For
example, the energy of a neutron in a constant background magnetic
field is given by
\begin{equation}
E = M_N - \mu_n B + 4\pi \, \frac{\beta_n}{2} B^2 + {\cal O}(B^3) \, .
\end{equation}
To extract both magnetic moments and magnetic polarizabilities, one
should select a field strength such that
\begin{equation}
\mu_n B \sim 4\pi \, \frac{\beta_n}{2} B^2 \,  .
\label{ideal}
\end{equation}
By considering both spin polarizations, the polarizability and the
magnetic moment may be isolated individually.  In order to use quark
propagators in the external field for both $u$ and $d$ quarks, the
magnetic field strengths should be selected in the ratio $-1:+2:-4:+8$
providing three different field strengths for study.  To satisfy the
criteria of (\ref{ideal}) at the intermediate field strength, one
requires a minimum field strength of
\begin{equation}
e \, B = { e \over 4\pi }\, {\mu_n \over \beta_n} \simeq 5\
\mbox{fm}^2 \, , 
\label{MinField}
\end{equation}
where $e^2 / 4 \pi = 1/137$ and experimental values of $\mu_n =
-1.913\ \mu_N$ and  $\beta_n = 0.3\, (10^{-3})$ fm${}^3$
\cite{schmiedmyer91} have been used for the estimate.  The
relationship for the smallest uniform magnetic field available on a
cubic lattice of cross-sectional area $A$ is \cite{smit87}
\begin{equation}
e\, B = {2\pi \over A \, q} \, ,
\end{equation}
where $q = 2/3$ or $-1/3$ for $u$ and $d$ quarks respectively.  To
accommodate $d$ quarks in the minimum uniform field of
(\ref{MinField}) requires an area $A \sim 4$ fm${}^2$.  We note that
with improved lattice QCD actions at lattice spacings of $\sim 0.25$
fm one requires a small $8^3$ lattice to satisfy this criteria.

   However the magnetic moment interaction energy associated with
(\ref{MinField}) is 200 MeV for the minimum field strength and grows
to 800 MeV.  Similarly the polarizability interaction energies range
from 100 MeV to 1600 MeV.  These values are too large to make contact
with the experimentally measured quantities.  Hence, one needs to
consider somewhat smaller field strengths.  Since the interaction
energy for polarizabilities varies as the square of the field
strength, only slightly smaller field strengths are required.  For
example, at lattice spacings of 0.23 fm, polarizability interaction
energies of 25, 100, and 400 MeV are provided by a $12^3$ lattice.
%
%
With modest supercomputer resources, these small lattices will provide
the opportunity to move beyond the quenched approximation and
investigate full QCD in which dynamical fermions are included.  Since
current lattice simulations use quark masses which are above the
physical quark masses, the remaining key source of systematic
uncertainty will lie in extrapolations to the physical quark masses.

   Hence, a determination of magnetic polarizabilities is readily
possible today.  It will be interesting to examine the effectiveness
of ${\cal O}(a^2)$-improved lattice QCD actions for describing
magnetic moments and polarizabilities.  Evaluations based on the
hadron spectrum are encouraging \cite{alford95D234,fiebig96,lee96a}.

   In summary, we have illustrated how Ioffe and Smilga's criticisms
\cite{ioffe84} of the lattice QCD calculations of Refs.\
\cite{bernard82,martinelli82} are ill-founded.  In fact, the
equivalence of the two approaches has been established.  It will be
interesting to reevaluate the utility of the background field approach
in light of the many advances made in lattice field theory since 1982.
Our hope is that clarification of these issues will generate a renewed
interest in the background field formalism as a viable method for the
extraction of hadron properties from QCD.

\acknowledgements

   D.L. thanks Kai Hencken for beneficial conversations.
M.B. acknowledges CEBAF and the Institute 
for Nuclear Theory at the University of Washington for their
hospitality and the U.S. Department of Energy for support during the
completion of this work.  D.B.L. acknowledges support from the
Natural Sciences and Engineering Research Council of Canada and the
U.S. Department of Energy under grant DE-FG06-88ER40427.
X.J. acknowledges support from the
Natural Sciences and Engineering Research Council of Canada.


\end{document}